\documentclass[preprint,showpacs,preprintnumbers,amsmath,amssymb,nofootinbib]{revtex4}

\usepackage[dvips]{graphicx}

\usepackage{graphicx}
\usepackage{dcolumn}
\usepackage{bm}
\usepackage{axodraw}

\newcommand{\bi}{\bibitem}
\newcommand{\be}{\begin{eqnarray}}
\newcommand{\ee}{\end{eqnarray}}
\newcommand{\nn}{\nonumber}
\catcode`\@=11
\def\lsim{\mathrel{\mathpalette\@versim<}}
\def\gsim{\mathrel{\mathpalette\@versim>}}
\def\@versim#1#2{\vcenter{\offinterlineskip
\ialign{$\m@th#1\hfil##\hfil$\crcr#2\crcr\sim\crcr } }}
\catcode`\@=12

\begin{document}
\def\descriptionlabel#1{\bf #1\hfill}
\def\description{\list{}{%
\labelwidth=\leftmargin
\advance \labelwidth by -\labelsep
\let \makelabel=\descriptionlabel}}

\vspace{2cm}
\preprint{KANAZAWA-06-13}

\title{$D_6$ Family Symmetry and Cold Dark Matter at LHC}

\author{Yuji Kajiyama}
\affiliation{National Institute of Chemical Physics and Biophysics,
Ravala 10, Tallinn 10143, Estonia
}
\author{Jisuke Kubo}
\author{Hiroshi Okada}

\affiliation{
Institute for Theoretical Physics, Kanazawa
University, Kanazawa 920-1192, Japan
\vspace{3cm}
}

\begin{abstract}
We consider a non-supersymmetric extension
of the standard model with a family symmetry based on
$D_6 \times Z_2 \times Z_2$, where one of $ Z_2$'s
is exactly conserved.
This $Z_2$  forbids the tree-level neutrino masses
and simultaneously ensures the stability of cold dark matter candidates.
From the assumption that cold dark matter is fermionic
we can single out the $D_6$ singlet right-handed neutrino
as the best cold dark mater candidate.
We find that an inert charged Higgs
with a mass between $300$ and $750$ GeV decays mostly into
an electron (or a positron) with a large missing energy, where
the missing energy is carried  away by the cold 
dark matter candidate.
This will be a clean signal at LHC.
\end{abstract}


\maketitle

\section{Introduction}
It is now clear that 
the standard model (SM) has to be extended 
at least in two ways:
Neutrino masses \cite{nmass} and 
cold dark matter  (CDM) \cite{wmap} have to be accommodated.
Since neutrinos are a part of dark matter \cite{wmap},
the nature of  cold dark matter and 
neutrinos may be somehow related.
A natural possibility to connect
apparently unrelated physics is  a symmetry.
By introducing an unbroken discrete symmetry, for instance,
we can make a weakly interacting neutral  particle
stable so that it can become a CDM candidate,
while making this discrete symmetry  responsible for
the smallness of neutrino masses.
Proposals along this line of thought have been 
suggested in refs. \cite{krauss,ma1,kubo1,hambye,kubo2}.
The basic idea of refs. \cite{krauss,ma1,kubo1} is to introduce
an unbroken $Z_2$ symmetry to forbid
tree-level neutrino masses and to assign an odd $Z_2$ parity
to  CDM candidates. For this mechanism
to work, one introduces an additional 
$SU(2)_L$ doublet Higgs, which does not acquire
VEV \cite{barbieri}, along with right-handed neutrinos.
We will adopt this idea in this paper.
However, the introduction of an additional
Higgs doublet and
the right-handed neutrinos into the SM introduces
additional ambiguities in the Yukawa sector.
Because of these ambiguities, 
it would be  difficult to make quantitative tests
of this idea, except may be the existence
of the additional  Higgs particles which could be found at LHC.

A natural guidance to constrain the Yukawa sector
is a flavor symmetry
\footnote{Recent flavor models are reviewed, for example,
in \cite{altarelli,mohapatra} and \cite{mondragon}}.
In this paper we would like to consider
a nonabelian discrete symmetry $D_6$,
which is one of the dihedral groups $D_N$.
The smallest dihedral group is $D_3$ which is isomorphic to
the smallest nonabelian
group $S_3$. $D_4$ has been used as a flavor symmetry 
in refs. \cite{seidl1,grimus2,hagedorn1},
while $D_5$ and $D_7$ have been considered 
in refs. \cite{hagedorn2} and \cite{chen}, respectively.
We will consider a non-supersymmetric extension of
the SM,
 which possesses a flavor symmetry based
 on $D_6 \times \hat{Z_2} \times Z_2$,
where $Z_2$ is exactly conserved and
$D_6 \times \hat{Z_2}$ is spontaneously broken
by the VEV of the $SU(2)_L$ doublet Higgs fields.
The unbroken $Z_2$ ensures the stability
of the CDM candidate and at the same time
forbids the tree-level neutrino masses,
while $\hat{Z_2}$ is responsible for the suppression of FCNCs
in the quark sector.
The $D_6$ assignment  is so chosen that
 the leptonic sector of the model is made predictive as possible
without having contradictions with experimental
observations in this sector:
There are eight independent parameters
to describe six lepton masses, and three mixing angles and three
CP violating phases of the neutrino mixing
matrix $V_{\rm MNS}$.
This will be discussed   in Sect. III,
where we will calculate the radiative
neutrino mass matrix \cite{zee}.

The $\mu\to e \gamma$ amplitude in our model does not
vanish \cite{ma3}. We will investigate it
in Sect. IV, making  it possible  to
single out the best  CDM candidate,
once we assume that CDM is fermionic. 
The fermionic CDM
candidate is the $D_6$ singlet right-handed
neutrino. Its mass  and the masses  of  
the additional Higgs fields are constrained 
by $\mu\to e \gamma$ and
the observed dark matter relic density.
In Sect. V we will plot these masses
from various aspects.
It will turn out,
among other things,  that 
the $D_6$ singlet inert charged Higgs
with a mass between $300$ and $750$ GeV
decays mostly into
an electron (or a positron) with a large missing energy, where
the missing energy is carried away by the CDM candidate.
They are within the accessible range of LHC \cite{zeppenfeld}.
Sect. V is devoted to summarizing our findings.

\section{The Model}

\subsection{$D_6$ group theory}
The dihedral groups, $D_N~(N=3,4,\dots)$,
are   nonabelian finite subgroups of $SO(3)$,
where 
all the irreps. of $D_N$ are real, and there exist
only two- and one-dimensional irreps \cite{frampton1,babu1}
\footnote{
The "covering group" of $D_N$ is $Q_{2N}$ \cite{kubo3},
which contains complex irreps. 
$Q_4$ and $Q_6$ have been considered in 
refs. \cite{frigerio} and \cite{frampton1,babu1}, respectively.}.
The irreps of $D_6$ are ${\bf 2}, 
 ~ {\bf2'},~{\bf 1},~{\bf 1'},~{\bf 1''},~{\bf 1'''}$, and
the group multiplication 
rules are given as follows \cite{frampton1}:
\begin{eqnarray}
{\bf 1'} \times {\bf 1'} &=& {\bf 1''} \times 
{\bf 1''} = {\bf 1'''} \times {\bf 1'''} = {\bf 1},~~
{\bf 1''} \times {\bf 1'''}= {\bf 1'},\nn\\
{\bf 1'} \times {\bf 1'''} &=& {\bf 1''},~~{\bf 1'} \times {\bf 1''} = {\bf 1'''}~.
\end{eqnarray}
The Clebsch-Gordan coefficients for multiplying  the  irreps
are  \cite{babu1}
\be
\begin{array}{ccccccccc}
 {\bf 2}  &  \times   
&  {\bf 2}  &  =  &  {\bf 1'} 
&  +   &  {\bf 1} & + &  {\bf 2'} 
\\ 
 \left(\begin{array}{c} x_1 \\ x_2  \end{array} \right)   & 
 \times    &  \left(\begin{array}{c} y_1 \\  y_2  \end{array} \right)  
&  =  &   (x_1 y_2 - x_2 y_1)   &  &
 (x_1 y_1 +x_2 y_2)   &    &
 \left(\begin{array}{c}x_1 y_1 - x_2 y_2 \\
 x_1 y_2 +x_2 y_1  \end{array} \right) ,\\ 
\end{array}\nn
  \label{multid61}
\ee
\be
\begin{array}{ccccccccc}
 {\bf 2'}  &  \times   
&  {\bf 2'}  &  =  &  {\bf 1} 
&  +   &  {\bf 1'} & + &  {\bf 2'} 
\\ 
 \left(\begin{array}{c}a_1 \\ a_2  \end{array} \right)   & 
 \times    &  \left(\begin{array}{c}b_1 \\  b_2   \end{array}\right)  
&  =  &   (a_1 b_1 + a_2 b_2)   &  &
 (a_1 b_2 -a_2 b_1)   &    &
 \left(\begin{array}{c}-a_1 b_1 + a_2 b_2 \\ a_1 b_2 +a_2 b_1 \end{array} \right) ,\\ 
\end{array}\nn
\label{multid62} 
\ee
\be
\begin{array}{ccccccccc}
 {\bf 2}  &  \times   
&  {\bf 2'}  &  =  &  {\bf 1'''} 
&  +   &  {\bf 1''} & + &  {\bf 2} 
\\ 
 \left(\begin{array}{c}x_1 \\ x_2  \end{array} \right)   & 
 \times    &  \left(\begin{array}{c}a_1 \\  a_2 \end{array} \right)  
&  =  &   (x_1 a_2 + x_2 a_1)   &  &
 (x_1 a_1 -x_2 a_2)   &    &
 \left(\begin{array}{c}x_1 a_1 + x_2 a_2 \\ x_1 a_2 -x_2 a_1 \end{array} \right) .  \\ 
\end{array}\nn
  \label{multid63}
\ee
In what follows we will use these multiplication rules 
to construct a flavor model.

\subsection{The Lepton Yukawa interaction}
The Yukawa sector of the SM contains a large number of
independent parameters.
We would like to reduce this number  to as few as possible
by a symmetry argument and 
then to understand
 the flavor structure in terms of the symmetry.
 In the following discussions we will concentrate on the
leptonic sector of a non-supersymmetric model.

As we will see below, it is possible to construct a model
with a family symmetry based on $D_6\times \hat{Z}_2\times Z_2$,
in which (1) the neutrino mass matrix ${\cal M}_\nu$ contains
only four real parameters, (2) the maximal mixing
of atmospheric neutrinos follows from the family symmetry,
and (3) the absolute value of
the $e-3$ element of the neutrino mixing matrix $V_{\rm MNS}$
can be expressed in terms of the charge lepton masses.
The  leptons $L, e^c, n$ and  $SU(2)_L$ Higgs doublets $\phi, \eta$
 belong to irreducible representations
of $D_6$, and we give  the 
$D_6\times \hat{Z}_2\times Z_2$ assignment in Table I and II.
$\hat{Z}_2\times Z_2$ is an abelian factor which
we impose on the model, where $Z_2$ shall remain unbroken after the
spontaneous symmetry breaking of  $SU(2)_L\times U(1)_Y$.
We use the two component notation for the Weyl spinors
in an obvious notation,
where $e^c$'s are the charge-conjugate
states of the right-handed electron family.
\begin{table}[thb]
\begin{center}
\begin{tabular}{|c|cccccc|} \hline
 & $L_S$ & $n_S$ & $e^c_S $&$L_I$&$n_I$&$e^c_I$ 
  \\ \hline
 $SU(2)_L\times U(1)_Y$ 
 & $({\bf 2}, -1/2)$  &  $({\bf 1}, 0)$  &  $({\bf 1}, 1)$
 & $({\bf 2}, -1/2)$&  $({\bf 1}, 0)$
 &  $({\bf 1}, 1)$
  \\ \hline
 $D_6$ & ${\bf 1}$  &  ${\bf 1}'''$  &  ${\bf 1}$
 & ${\bf 2}'$&  ${\bf 2}'$&  ${\bf 2}'$
 \\ \hline
 $\hat{Z}_2$
 & $+$ &$+$  & $-$  & $+$ 
 &$+$  & $-$
  \\ \hline
   $Z_2$
 & $+$ &$-$  & $+$  & $+$ &$-$  & $+$ 
   \\ \hline
\end{tabular}
\caption{The $D_6 \times \hat{Z}_2\times Z_2$ 
assignment for the leptons.  The subscript $S$ indicates
a $D_6$ singlet, and the subscript $I$ running from $1$ to $2$
stands for a $D_6$ doublet. $L$'s denote the $SU(2)_L$-doublet leptons,
while $e^c$ and $n$ are the $SU(2)_L$-singlet leptons.
}
\end{center}
\end{table}
\begin{table}[thb]
\begin{center}
\begin{tabular}{|c|cccc|} \hline
 & $\phi_S$ &$\phi_I$ & $\eta_S$&$\eta_I $
  \\ \hline
   $SU(2)_L\times U(1)_Y$ 
 & $({\bf 2}, -1/2)$  &  $({\bf 2}, -1/2)$   &  $({\bf 2}, -1/2)$ 
 & $({\bf 2}, -1/2)$ 
  \\ \hline
 $D_6$ & ${\bf 1}$ &${\bf 2}'$  &  ${\bf 1}'''$  & ${\bf 2}'$
 \\ \hline
 $\hat{Z}_2$ &$+$ 
 & $-$ &$+$  & $+$ 
  \\ \hline
   $Z_2$
 & $+$ &$+$  & $-$  & $-$ 
   \\ \hline
\end{tabular}
\caption{The $D_6 \times \hat{Z}_2\times Z_2$ 
assignment  for the $SU(2)_L$  Higgs doublets.
}
\end{center}
\end{table}
Under  $Z_2$  (which plays the rolle of $R$ parity
in the MSSM), only the right-handed neutrinos $n_S, n_I$ and 
the extra Higgs $\eta_S, \eta_I$ are odd.
The quarks are assumed to belong to ${\bf 1}$ of $D_6$ 
with $(+,+)$ of  $\hat{Z}_2\times Z_2$ so that the 
quark sector is basically the same as the SM, where 
the $D_6$ singlet Higgs $\phi_S$ with
$(+,+)$ of  $\hat{Z}_2\times Z_2$ plays the rolle of the SM Higgs
in this sector. No other Higgs can couple to the quark sector at the tree-level.
In this way we can avoid tree-level FCNCs in the quark sector.
So, $\hat{Z}_2$ is introduced to forbid  tree-level couplings of 
the $D_6$ singlet Higgs $\phi_S$ with the leptons
and simultaneously to forbid  tree-level couplings of $\phi_I, \eta_I$ and $\eta_S$
with the quarks.

The most general renormalizable $D_6 \times \hat{Z}_2 \times 
Z_2$ invariant 
Yukawa interactions in the leptonic sector 
can be described by
\be
{\cal L}_{Y} &=&\sum_{a,b,d=1,2,S}~\left[
Y_{ab}^{ed} (L_{a} i\sigma_2\phi_d) e^c_{b} 
+Y_{ab}^{\nu d} (\eta_d^\dag L_{a}) n_{b} \right]
- \sum_{I=1,2}\frac{1}{2}M_{1} n_{I}n_{I}-
 \frac{1}{2}M_{S}n_{S}n_{S}+h.c.
 \label{wL}
\ee
The Yukawa matrices $Y$'s are given by
\be
{\bf Y}^{e1} &=&\left(\begin{array}{ccc}
-y_2 & 0 & y_5\\
0 & y_2 &  0\\
y_4 & 0  & 0 \\
\end{array}\right),~
{\bf Y}^{e2} =\left(\begin{array}{ccc}
0 & y_2 & 0 \\
y_2 & 0 & y_5 \\
0&  y_4 & 0 \\
\end{array}\right),~
{\bf Y}^{eS}=0,
\label{Yue}
\ee
\be
{\bf Y}^{\nu1} &=&\left(\begin{array}{ccc}
-h_2& 0 & 0 \\
0 & h_2& 0 \\
h_4 & 0 & 0 \\
\end{array}\right),~
{\bf Y}^{\nu2} =\left(\begin{array}{ccc}
0 & h_2 & 0\\
h_2&  & 0 \\
0&h_4 & 0 \\
\end{array}\right),\nn\\
{\bf Y}^{\nu S}&=&\left(\begin{array}{ccc}
0 & 0 & 0\\
0 & 0 & 0 \\
0 &  0 &h_3 \\
\end{array}\right).
\label{Yun}
\ee

\subsection{The scalar potential and the $\eta$ masses}
The most general renormalizable 
Higgs potential, invariant under
$D_6 \times \hat{Z}_2\times Z_2$,  is given by
\footnote{See also  for instance \cite{okada1}.}
\be
V (\phi,\eta)&=&V_1[\phi;\mu_1^\phi,\mu_2^\phi,
\lambda_1^\phi,\dots,\lambda_7^\phi]
+V_1[\eta;\mu_1^\eta,\mu_2^\eta,
\lambda_1^\eta,\dots,\lambda_7^\eta ]+
V_2[\phi,\eta],
\label{pot0}
\ee
where
\be
& &V_1[\phi;\mu_1,\mu_2,
\lambda_1,\dots, \lambda_7]\nn\\
& &=-\mu_1^2(\phi_I^\dag \phi_I)
-\mu_2^2(\phi_S^\dag \phi_S)
 +V_3[(\phi^\dag \phi),
(\phi^\dag \phi); \lambda_1, \lambda_2,\lambda_3] \nn\\
& &+\lambda_4(\phi_S^\dag \phi_S)(\phi_I^\dag \phi_I)+
\lambda_5(\phi_S^\dag \phi_I)(\phi_I^\dag \phi_S)+
[\lambda_6(\phi_S^\dag \phi_I)^2+h.c.]+
\lambda_7(\phi_S^\dag \phi_S)^2,
\label{pot1}\\
& &V_2[\phi,\eta]\nn\\
& &=
V_3[(\phi^\dag \phi),(\eta^\dag \eta); \kappa_1,\kappa_2,\kappa_3]\nn\\
& &+\kappa_4(\phi^\dag_S \phi_S)(\eta^\dag_I \eta_I)
+\kappa_5(\phi^\dag_I \phi_I)(\eta^\dag_S \eta_S)
+\kappa_6(\phi^\dag_S \phi_S)(\eta^\dag_S \eta_S)\nn\\
& &+ V_3[(\phi^\dag \eta),(\eta^\dag \phi); \kappa_7,
\kappa_8,\kappa_9]\nn\\
& &+\kappa_{10}(\phi^\dag_S \eta_I)(\eta^\dag_I \phi_S)
+\kappa_{11}(\phi^\dag_I \eta_S)(\eta^\dag_S \phi_I)
+\kappa_{12}(\phi^\dag_S \eta_S)(\eta^\dag_S\phi_S)\nn\\
& &+\left\{~ V_3[(\phi^\dag \eta),(\phi^\dag \eta); 
\kappa_{13},\kappa_{14},\kappa_{15}]\right.\nn\\
& &\left.+\kappa_{16}(\phi^\dag_S \eta_I)(\phi^\dag_S \eta_I)
+\kappa_{17}(\phi^\dag_I \eta_S)(\phi^\dag_I \eta_S)
+\kappa_{18}(\phi^\dag_S \eta_S)(\phi^\dag_S \eta_S)
+h.c.~\right\},
\label{pot2}
\ee
and $I$ runs from $1$ to $2$.
Here $V_3$ is defined as
\be
& &V_3[(A^\dag B),(C^\dag D); \kappa_1,\kappa_2,\kappa_3] \nn\\
&=&\kappa_1(A_I^\dag  B_I)(C_J^\dag  D_J)+
\kappa_2(A_I^\dag  (i\sigma_2)_{IJ} B_J)
(C_K^\dag (i\sigma_2)_{IJ} D_L)\nn\\
& &+\kappa_3
[~(A_I^\dag  (\sigma_1)_{IJ} B_J)(C_K^\dag (\sigma_1)_{IJ} D_L)+
(A_I^\dag  (\sigma_3)_{IJ} B_J)(C_K^\dag (\sigma_3)_{IJ} D_L)~],
\label{pot3}
\ee
where $A, B, C$ and $D$ are $SU(2)_L$ doublets
and belong to ${\bf 2}' $ of $D_6$, 
and $(A^\dag B)$ is  an $SU(2)_L$ invariant product.
As we can see from (\ref{pot2}) and (\ref{pot3}),  
an exact lepton number $U(1)_L'$ invariance emerges
in the absence of $\kappa_{13} \sim \kappa_{18}$,
where the right-handed neutrinos $n_I$ and $ n_S$
are neutral under $U(1)_L'$ in contrast to
the conventional seesaw models \cite{seesaw}.
This $U(1)_L'$ forbids the neutrino masses,
so that the smallness of the neutrino masses
has a natural meaning.
Therefore, the radiative neutrino masses will be
proportional to these Higgs couplings.

We assume that $(\mu_1^\eta)^2, (\mu_2^\eta)^2 < 0$ so that 
\be
<\eta_S>=<\eta_I>=0
\label{VEVeta}
\ee
 corresponds to a local minimum
of the scalar potential  (\ref{pot0}). This is the essence of
 refs. \cite{ma1,kubo1}
to connect the neutrino masses 
and the nature of CDM,
because $Z_2$ remains unbroken.
We further observe that the scalar potential (\ref{pot0})
 has an accidental symmetry $S_2$:
\be
\phi_{1}, \eta_1 &\leftrightarrow & \phi_{2},\eta_2,
\label{s2p}
\ee
while the $D_6$ singlets Higgs fields
$\phi_S$ and $\eta_S$ remain unchanged.
This symmetry ensures that if
$(\mu_1^\phi)^2, (\mu_2^\phi)^2 > 0$,
\be
<\phi_1>=<\phi_2>&=&
\left(\begin{array}{c}
 v_D/2 \\ 0
\end{array}\right)
~~\mbox{and}~~<\phi_S>=
\left(\begin{array}{c}
 v_S/\sqrt{2}\\
0
\end{array}\right)
\label{VEVphi}
\ee
can correspond to a local minimum of the potential, suggesting
an appropriate field redefinition
\footnote{
The tree-level W boson mass constraint is 
$(v_D^2+v_S^2)=v^2\simeq (246\,\mbox{GeV})^2$.}
\be
\phi_\pm &=&\frac{1}{\sqrt{2}}(\phi_1\pm\phi_2)~,~
\eta_\pm =\frac{1}{\sqrt{2}}(\eta_1\pm\eta_2).
\label{phipm}
\ee

A consequence of (\ref{VEVeta}) is that $\eta'$s do not mix with $\phi'$s
in the mass matrix. Furthermore, because of the absence
of the $\phi_S \phi_I\eta_S \eta_I$ type couplings,
$\eta_\pm$ and $\eta_S$ dot not mix with each other.
 Keeping these observations in mind, we 
can write down the mass terms for $\eta$'s as
\be
{\cal L}_{M_\eta} &=&
-\sum_{a=+,-,S}\left[~m_a^2\eta_a^{(+)}\eta_a^{(-)}
+\frac{1}{2}\left(m_a^R\right)^2\eta_a^{(0)}\eta_a^{(0)}
+\frac{1}{2}\left(m_a^I\right)^2\chi_a^{(0)}\chi_a^{(0)}~\right.\nn\\
& &\left. +\left(m_a^{RI}\right)^2\eta_a^{(0)}\chi_a^{(0)}~\right],
\label{lmeta}
\ee
where $\eta^{(-)}
~(=(\eta^{(+)})^*),\eta^{(0)}$ and $\chi^{(0)}$ are
$SU(2)_L$ components of $\eta$, i.e.,
\be
\eta &=&\left(\begin{array}{c}
 (\eta^{(0)}+i \chi^{(0)})/\sqrt{2} \\
 \eta^{(-)} 
\end{array}\right).
\label{eta0}
\ee
Since the neutrino masses will be proportional to
the $U(1)_L'$ violating Higgs couplings
$\kappa_{13}\sim \kappa_{18}$, we may assume 
that they are small.
In the absence of these Higgs couplings,
there will be no mixing of the scalar and 
pseudo scalar components of
the neutral $\eta'$s, i.e., $m_a^{RI}=0$, and $m_a^{R}=m_a^{I}$.
In general, even though it is small, there exists mixing:
\be
\eta^{(0)}_a  &=& \cos\gamma_a \hat{\eta}_a
 +\sin\gamma_a \hat{\chi}_a~,~
 \chi^{(0)}_a  =-\sin\gamma_a \hat{\eta}_a
 +\cos\gamma_a \hat{\chi}_a,~~(a=\pm,S),
 \label{gammaa}
\ee
where $\hat{\eta}_a$ and $\hat{\chi}_a$ are mass eigenstates.
We denote their masses by
$m_a^\eta$ and $m_a^\chi$, respectively.
The difference $\left(m_a^\eta\right)^2- \left(m_a^\chi\right)^2$
will be proportional to the $U(1)'_L$ violating
couplings, and we find that
\be
\Delta m_a^2 &=&
\left(m_a^\eta\right)^2- \left(m_a^\chi\right)^2\nn\\
&=&[ (m^R_a)^2-(m^I_a)^2 )]\cos2
\gamma_a-2(m^{RI}_a)^2\sin2\gamma_a,
\label{deltama}
\ee
where
\be
 (m^R_a)^2-(m^I_a)^2 
 &=&\left\{~\begin{array}{c}
2[(\mbox{Re}v_D)^2-
(\mbox{Im}v_D)^2](\mbox{Re}\kappa_{13}+\mbox{Re}\kappa_{15})
+2[(\mbox{Re}v_S)^2-(\mbox{Im}v_S)^2]\mbox{Re}\kappa_{16}\\
+4\mbox{Re}v_D
\mbox{Im}v_D(\mbox{Im}\kappa_{13}+\mbox{Im}\kappa_{15})
+4\mbox{Re}v_S\mbox{Im}v_S\mbox{Im}\kappa_{16}\\
\dotfill\\
2[(\mbox{Re}v_D)^2-
(\mbox{Im}v_D)^2](\mbox{Re}\kappa_{14}+\mbox{Re}\kappa_{15})
+2[(\mbox{Re}v_S)^2-(\mbox{Im}v_S)^2]\mbox{Re}\kappa_{16}\\
+4\mbox{Re}v_D
\mbox{Im}v_D(\mbox{Im}\kappa_{14}+\mbox{Im}\kappa_{15})
+4\mbox{Re}v_S\mbox{Im}v_S\mbox{Im}\kappa_{16}\\
\dotfill\\
2[(\mbox{Re}v_D)^2-(\mbox{Im}v_D)^2]Re\kappa_{17}
+4\mbox{Re}v_D \mbox{Im}v_D\mbox{Im}\kappa_{17}\\
+
2[(\mbox{Re}v_S)^2-(\mbox{Im}v_S)^2] \mbox{Re}\kappa_{18}
+ 4\mbox{Re}v_S \mbox{Im}v_S \mbox{Im}\kappa_{18}
\end{array} \right. 
\nn\\
(m^{RI}_a)^2 &=&\left\{~\begin{array}{c}
2\mbox{Re}v_D\mbox{Im}v_D(\mbox{Re}\kappa_{13}
+\mbox{Re}\kappa_{15})
+2\mbox{Re}v_S\mbox{Im}v_S\mbox{Re}\kappa_{16}\\
-[(\mbox{Re}v_D)^2-(\mbox{Im}v_D)^2]
(\mbox{Im}\kappa_{13}+\mbox{Im}\kappa_{15})
-[(\mbox{Re}v_S)^2-(\mbox{Im}v_S)^2]\mbox{Im}\kappa_{16}
\\  
\dotfill\\
2\mbox{Re}v_D\mbox{Im}v_D(\mbox{Re}\kappa_{14}
+\mbox{Re}\kappa_{15})
+2\mbox{Re}v_S\mbox{Im}v_S\mbox{Re}\kappa_{16}\\
-[(\mbox{Re}v_D)^2-(\mbox{Im}v_D)^2]
(\mbox{Im}\kappa_{14}+\mbox{Im}\kappa_{15})
-[(\mbox{Re}v_S)^2-(\mbox{Im}v_S)^2]\mbox{Im}\kappa_{16}\\ 
\dotfill\\
2\mbox{Re}v_D\mbox{Im}v_D\mbox{Re}\kappa_{17}+
2\mbox{Re}v_S\mbox{Im}v_S\mbox{Re}\kappa_{18}\\
-[(\mbox{Re}v_D)^2-(\mbox{Im}v_D)^2]\mbox{Im}\kappa_{17}
-[(\mbox{Re}v_S)^2-(\mbox{Im}v_S)^2]\mbox{Im}\kappa_{18}
\end{array} \right.\nn
 \ee
for $a=+, -, S$.
As we will see below,
the absence of the mixing among
 $\eta_\pm$ and 
$\eta_S$ is responsible for the reason that
the neutrino masses
and mixing
of the present model have the same structure
as that of the $S_3$ model of ref. \cite{kubo4}.

\section{Lepton masses and mixing}
\subsection{CP phases}
 Let us first figure out the structure of CP phases.
To this end, we
introduce phases explicitly as follows:
\be
y_{a}&\to &e^{i p_{y_{a}}} y_{a} ~(a=2,4,5)
\label{phases}
\ee
for the Yukawa couplings, 
where we assume that the possible phases coming from the VEVs of
$\phi'$s are absolved into the  Yukawa couplings.
The  $y$'s
on the right-hand side are supposed to be real and
$-\pi/2 \leq p's\leq \pi/2$, and similarly for the fields
\be
L_{I}&\to & e^{i p_{L}}L_{I}~,~
L_{S} \to e^{i p_{L_{S}}}L_{S}~,~
e_{I}^c\to  e^{i p_{e}}e_{I}^c~,~
e_{S}^c \to e^{i p_{e_{S}}}e_{S}^c,\nn\\
n_{I} &\to &  e^{i p_{n}}n_{I}~,~
n_{S} \to e^{i p_{n_{S}}}n_{S}.
\ee
The phases of the right-handed neutrinos 
are used to absorb the phases of their Majorana
masses $M_{1}$ and $M_{S}$.
The phases of $y_{2},y_{4}$ and $y_{5}$ can be rotated away if
\be
0&=&p_{L}+  p_{y_{2}} +p_{e}~,~
0=p_{L_{S}} +p_{y_{4}}+p_{e}~,
~0=p_{L} +p_{y_{5}}+p_{e_S}
\label{pL}
\ee
are satisfied. So, only one free  phase is left,
which we assume to be $p_{L}$.
We will use this freedom to make certain 
entries of the one-loop neutrino mass matrix real.
After that no further phase rotation 
which does not change physics is possible.

\subsection{Charged fermion masses}
The charged lepton masses are generated from the $S_{2}$
invariant VEVs (\ref{VEVphi}), and the mass matrix becomes
\be
{\bf M}_{e} = \left( \begin{array}{ccc}
-m_{2} & m_{2} & m_{5} 
\\  m_{2} & m_{2} &m_{5}
  \\ m_{4} & m_{4}&  0
\end{array}\right),
\label{mlepton}
\ee
where
\be
m_{2} &=& | v_D y_{2} |
/2, m_{4}= |v_D  y_{4}| /2,
m_{5}= |v_D y_{5}|/2.
\label{m2-m5}
\ee
All the mass parameters
appearing in (\ref{mlepton}) can be assumed to be real.
Diagonalization of the mass matrices is straightforward.
The mass eigen values are
approximately given by \cite{kubo4}
\be
m_e^2 &=& \frac{(m_{4} 
m_{5})^2}{(m_{2})^2+(m_{5})^2}
+O((m_{4})^4),\\
~m_\mu^2 &=& 2 (m_{2})^2+
(m_{4})^2+O((m_{4})^4),\\
m_{\tau}^2 &=&  2[~(m_{2})^2+(m_{5})^2~]+
\frac{(m_{4} m_2)^2}{(m_2)^2
+(m_{5})^2}+O((m_{4})^4).
\ee
Concrete values are given as
$m_4/m_5\simeq 0.00041$ and $ m_2/m_5 \simeq 0.0596$ and
$m_{5}\simeq  1254$ MeV
to obtain 
$m_e=0.51$ MeV, $m_\mu =105.7$ MeV and $m_\tau
=1777$ MeV.
The diagonalizing unitary matrices
(i.e., $U_{eL}^{T} {\bf M}_{e} U_{eR}$)
assume a simple form in the $m_{e} \to 0$ limit,
which is equivalent to the $m_{4} \to 0$ limit.
We find that $U_{eL}$ can be approximately written as
\cite{kobayashi1}
\be
U_{eL} &= &\left(
\begin{array}{ccc}
\epsilon_e( 1 - \epsilon_\mu^2) &
-(1/\sqrt{2}) (1-\epsilon_e^2+2\epsilon_e^2 \epsilon_\mu^2) &
1/\sqrt{2}\\
-\epsilon_e( 1 + \epsilon_\mu^2) &
(1/\sqrt{2})(1-\epsilon_e^2-2\epsilon_e^2 \epsilon_\mu^2 ) &
1/\sqrt{2} \\
1-\epsilon_e^2
& \sqrt{2}\epsilon_e &  \sqrt{2} \epsilon_e \epsilon_\mu^2
\end{array}\right),
\label{UeL}
\ee
where $\epsilon_\mu =m_{\mu}/ m_{\tau}$ and
$\epsilon_e =m_e/(\sqrt{2}m_{\mu})$.
In the limit $m_e=0$, the unitary matrix $U_{eL}$
becomes
\be
\left(
\begin{array}{ccc}
0&-1/\sqrt{2} &1/\sqrt{2}\\
0 &1/\sqrt{2}  &1/\sqrt{2}\\1& 0 &  0
\end{array}\right),
\ee
which is the origin of a maximal mixing of
the atmospheric neutrinos.

\subsection{Radiative neutrino masses}
\begin{figure}[htb]
\begin{center}\begin{picture}(300,110)(0,45)
\ArrowLine(70,50)(110,50)
\Line(110,50)(190,50)
\ArrowLine(230,50)(190,50)
\Text(90,35)[b]{$\nu_L$}
\Text(210,35)[b]{$\nu_L$}
\Text(150,35)[b]{$N$}
\Text(150,100)[]{$\eta^{(0)},\chi^{(0)}$}
\Text(150,50)[]{$\times $}
\DashCArc(150,50)(40,0,180){3}
\end{picture}\end{center}
\caption[]{\label{numass}One-loop radiative neutrino mass.}
\end{figure}
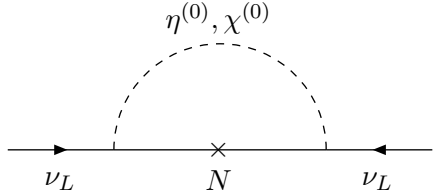
The neutrino mass matrix is generated from the one-loop 
diagram fig. \ref{numass} \cite{ma1} and 
is given by 
\be
({\cal M}_\nu)_{ij} &=& 
\sum_{a=\pm,S}\sum_{k=1,2,S} 
({\bf Y}^{\nu a})_{ik}^*
( {\bf Y}^{\nu a})_{jk}^* \Gamma^{a}(M_k),
\label{mnu}
\ee
where
\be
 {\bf Y}^{\nu +} &=&\frac{1}{\sqrt{2}}
 (~{\bf Y}^{\nu 1}+{\bf Y}^{\nu 2}~)=
 \frac{1}{\sqrt{2}}
 \left(\begin{array}{ccc}
 -h_2 & h_2 &0\\
  h_2 & h_2 &0\\
 h_4 & h_4 &0
 \end{array}\right),
 \label{ynup}\\
  {\bf Y}^{\nu -} &=&\frac{1}{\sqrt{2}}
 (~{\bf Y}^{\nu 1}-{\bf Y}^{\nu 2}~)=
 \frac{1}{\sqrt{2}}
 \left(\begin{array}{ccc}
 -h_2 &- h_2 &0\\
  -h_2 & h_2 &0\\
 h_4 & -h_4 &0
 \end{array}\right),
 \label{ynum}\\
\Gamma^{a}(M_k) &=&\frac{M_k}{8\pi^2} 
\exp(-i2\gamma_a)\left[
\frac{(m_a^\eta/M_k)^2 \ln (m_a^\eta/M_k)^2}{1-(m_a^\eta/M_k)^2}-
\frac{(m_a^\chi/M_k)^2 
\ln (m_a^\chi/M_k)^2}{1-(m_a^\chi/M_k)^2}\right]\\
& \simeq &\frac{M_k}{8\pi^2} \exp(-i2\gamma_a)\Delta m_a^2 
\frac{1-(m_a^\eta/M_k)^2+
\ln (m_a^\eta/M_k)^2}{(1-(m_a^\eta/M_k)^2)^2}.
\label{func-i}
\ee
The Yukawa matrices, $m_a^{\eta,\chi}$,
$\gamma_a$ and $\Delta m_a^2$ 
are defined in 
(\ref{Yue}),
(\ref{Yun}), (\ref{lmeta}), (\ref{gammaa}) and (\ref{deltama}), respectively.
(Recall that $M_1=M_2$ because of the $D_6$ symmetry.)

Using the explicit form of the Yukawa matrices we obtain
\be
{\cal M}_\nu &=&
\left(\begin{array}{ccc}
G^+ (M_1)h_2^2 & 0 & 0\\
0 & G^+ (M_1)h_2^2 &
G^- (M_1)h_2 h_4\\
0 & G^- (M_1)h_2 h_4 &
G^+ (M_1)h_4^2+\Gamma^S(M_S)h_3^2
\end{array}\right),
\label{mnu1}
\ee
where
\be
G^\pm (M_1)&=& \Gamma^+(M_1)\pm\Gamma^-(M_1).
\ee
At this stage we recall that the phase $p_L$
was left
undetermined (see  (\ref{pL}) ), and so we use it to make the $(1,1)$
and $(2,2)$ entries of ${\cal M}_\nu$ real.
Then $(2,3)$
and $(3,2)$ entries of ${\cal M}_\nu$ can be made real by
multiplying a diagonal phase matrix
\be
P &=&\left(\begin{array}{ccc}
1 & 0 & 0\\
0 & 1 & 0\\
0 & 0& \exp( i p)
\end{array}\right)
\label{phase}
\ee
with ${\cal M}_\nu$ from left and right, and we can rewite
the neutrino mass matrix as
\be
P {\cal M}_\nu P &=& {\bf M}_{\nu} = \left( \begin{array}{ccc}
2 (\rho_{2})^2 & 0 & 
0
\\ 0 & 2 (\rho_{2})^2 & 2 \rho_2 \rho_{4}
  \\ 0 & 2 \rho_2 \rho_{4}  &  
2 (\rho_{4})^2 +
(\rho_3)^2\exp i 2 \varphi_{3}
\end{array}\right),
\label{m-nu}
\ee
where $p$ is an independent parameter and enters
into the definition
of the  neutrino mixing matrix $V_{MNS}$, and
the $\rho$'s in (\ref{m-nu}) are real numbers.
One can convince oneself that ${\bf M}_{\nu}$ 
can be diagonalized as \cite{kubo4,mondragon}
\be
U^T_\nu {\bf M}_{\nu} U_\nu &=& \left( \begin{array}{ccc}
m_{\nu_1}e^{i\phi_1-i\phi_\nu} & 0 & 0\\
0 & m_{\nu_2}e^{i\phi_2+i\phi_\nu} &0 \\
0 & 0 & m_{\nu_3}
\end{array}\right),
\ee
where
\be
U_{\nu}&= &\left( \begin{array}{ccc}
 0 & 0 &1\\
 -s_{12} & c_{12}e^{i \phi_\nu}
&  0\\
    c_{12}e^{-i \phi_\nu}  & s_{12}& 0
 \end{array}\right),\\
\label{unumax3}
m_{\nu_3} \sin \phi_\nu &=& m_{\nu_2} \sin \phi_2
=m_{\nu_1} \sin \phi_1~,~2 \varphi_{3}=\phi_{1}+\phi_{2}\\
\label{sinp}
m_{\nu_3} &=& 2 \rho_{2}^{2},~
\frac{m_{\nu_1}m_{\nu_2}}{m_{\nu_3}}= \rho_{3}^{2},
\label{rho23}\\
\tan \phi_\nu &=&
\frac{\rho_3^2 \sin2\varphi_3}{2(\rho_2^2+\rho_4^2)
+\rho_3^2\cos2\varphi_3},
\label{tanvarphi}
\ee
and $c_{12}=\cos\theta_{12}$ and $s_{12}=\sin\theta_{12}$.
We also find that
\be
\tan^2\theta_{12} &=&
\frac{(m_{\nu_2}^2-m_{\nu_3}^2 \sin^2\phi_\nu)^{1/2}
-m_{\nu_3}|\cos\phi_\nu|}{(m_{\nu_1}^2
-m_{\nu_3}^2 \sin^2\phi_\nu)^{1/2}
+m_{\nu_3}|\cos\phi_\nu|},
\ee
from which we obtain
\be
\frac{m_{\nu_2}^2}{\Delta m_{23}^2} &=&
\frac{(1+2 t_{12}^2+t_{12}^4-r t_{12}^4)^2}
{4  t_{12}^2 (1+t_{12}^2)(1+t_{12}^2-r t_{12}^2)\cos^2 \phi_\nu}
-\tan^2 \phi_\nu
\label{mnu2}\\
&\simeq &
\frac{1}{\sin^2 2\theta_{12}\cos^2 \phi_\nu}
-\tan^2 \phi_\nu ~~\mbox{for}~~|r| << 1,
\label{mnu21}
\ee
where $t_{12}=\tan\theta_{12}, r=\Delta m_{21}^2/\Delta m_{23}^2$.
It can also be shown  that  only an inverted mass spectrum
\be
m_{\nu_3} & < & m_{\nu_1}, m_{\nu_2}
\label{spectrum}
\ee
is consistent with  the experimental constraint $ |\Delta m_{21}^2|
< |\Delta m_{23}^2|$  in the present model.
Note that eq. (\ref{sinp}) is satisfied for
\be
2 \varphi_{3} &=& \phi_{1}+\phi_{2} \sim \pm\pi
\label{1plus2}
\ee
and not for $\phi_{1} \sim \phi_{2}$. That is,
if $2 \varphi_{3}  \sim +(-)\pi$, then
$\cos\phi_1 < (>) 0$ and $\cos\phi_2 > (<) 0$.

Now the product $U_{eL}^{\dag}P  U_\nu$ 
defines the neutrino mixing matrix $V_{\rm MNS} $,
where $P$ is defined in (\ref{phase}). We find
\be
|(V_{\rm MNS})_{13}| &=&\frac{m_e}{\sqrt{2}m_\mu}
+O(m_e m_\mu/m_\tau^2)
\simeq 3.4\times 10^{-3},\\
|(V_{\rm MNS})_{23}| &=&\frac{1}{\sqrt{2}}+O((m_e /m_\mu)^2),~
|(V_{\rm MNS})_{33}| =\frac{1}{\sqrt{2}}.
\ee
So the mixing of the atmospheric neutrinos are very close
to the maximal form 
\footnote{
Unfortunately, this value of $|(V_{\rm MNS})_{13}| $ is too small
to be measured \cite{minakata}.}.
A reparametrization independent
indicator for CP violation in neutrino oscillations is 
the Jalskog determinant, which is given by
\be
|J| &=&\left|\mbox{Im}~[(V_{\rm MNS})_{22}(V_{\rm MNS})_{33}
(V_{\rm MNS}^*)_{23}(V_{\rm MNS}^*)_{32}]\right|\nn\\
&=&
\frac{m_e}{\sqrt{2}m_\mu}\sin 2(\theta_{12})\sin(p-\phi_\nu)
+O(m_e m_\mu/m_\tau^2) < 3.5\times 10^{-4}
\ee
in the present model.

\begin{center}
\begin{figure}[htb]
\includegraphics*[width=0.6\textwidth]{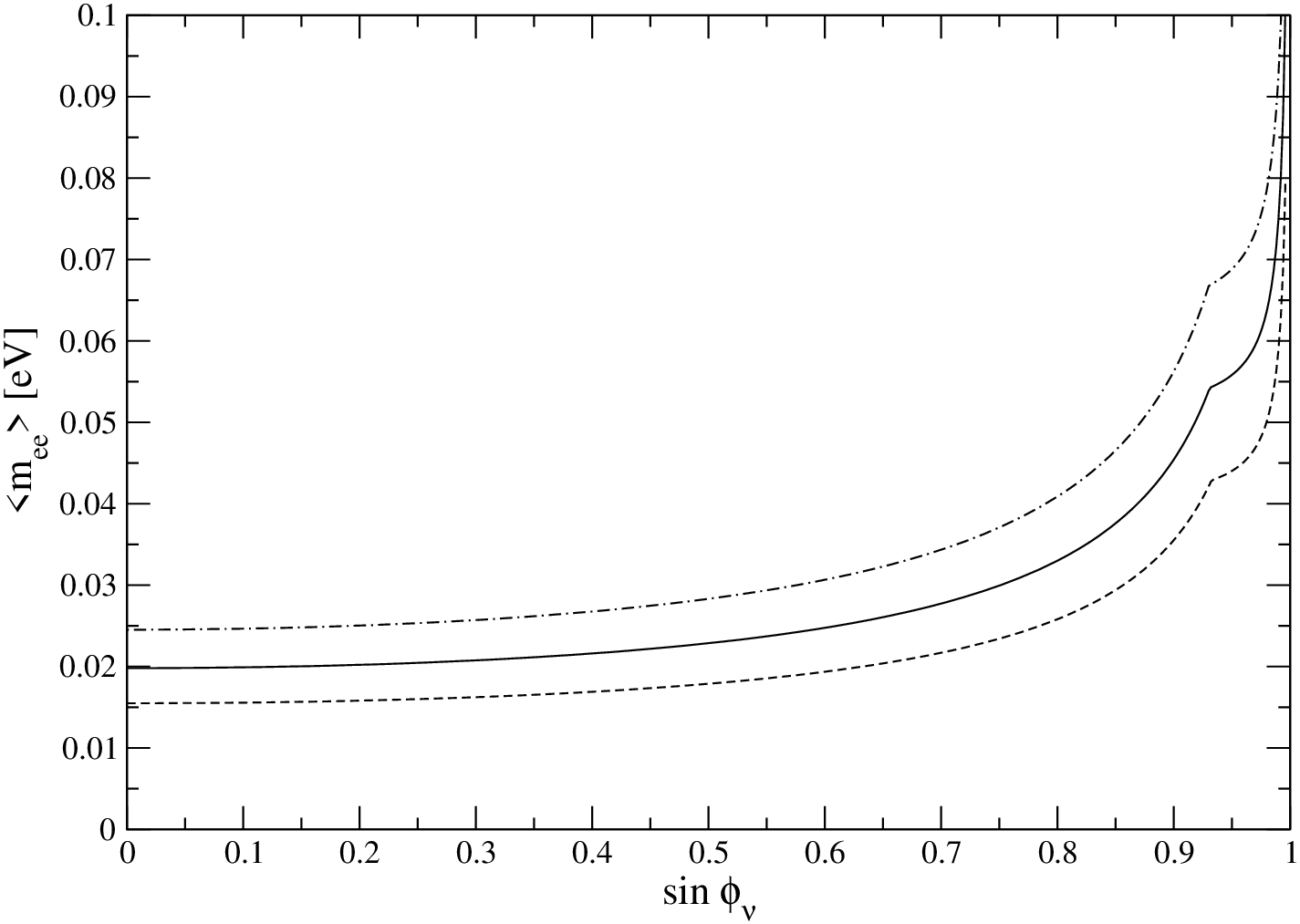}
\caption{\label{mee}
The effective Majorana mass $<m_{ee}>$ as a function of
$\sin \phi_\nu$ with 
$\sin^2\theta_{12}=0.3$ and 
 $\Delta m_{21}^2=6.9 \times10^{-5}$ eV$^2$.
 The dashed, solid and dot-dashed lines stand for 
$\Delta m_{23}^2=1.4, 2.3$ and $ 3.0 \times 10^{-3}$ eV$^2$,
respectively. The $\Delta m_{21}^2$ dependence is very small.}
\end{figure}
\end{center}

The effective Majorana mass $<m_{ee}>$ in neutrinoless
double $\beta$ decay is  given by
\be
<m_{ee}> &=& |~\sum_{i=1}^3m_{\nu_i} V_{ei}^2|
\simeq |m_{\nu_1}c_{12}^2+
m_{\nu_2}s_{12}^2 \exp i2\alpha~ |,
\label{mee1}
\ee
where
\be
\sin 2 \alpha &=&\sin(\phi_1-\phi_2)\nn\\
& =&
\pm \frac{ m_{\nu_3}\sin\phi_\nu}{m_{\nu_1}m_{\nu_2}}
\left( \sqrt{m_{\nu_2}^2-m_{\nu_3}^2 \sin^2 \phi_\nu}+
\sqrt{m_{\nu_1}^2-m_{\nu_3}^2 \sin^2 \phi_\nu} \right)
\label{alpha}
\ee
for $2\varphi_2 \sim \pm \pi$, and
$\phi_1,\phi_2$ and $\phi_\nu$ are defined in (\ref{sinp}).
In fig.~2 we  plot $<m_{ee}>$  as a function of $\sin \phi_\nu$
for $\sin^2\theta_{12}=0.3, \Delta m_{21}^2=6.9 \times10^{-5}$ eV$^2$ and
$\Delta m_{23}^2=1.4, 2.3, 3.0 \times 10^{-3}$ eV$^2$ \cite{maltoni3}.
As we can see from fig.~\ref{mee}, the effective Majorana mass stays
at about its minimal value $<m_{ee}>_{\rm min}$ 
for a wide range of $\sin\phi_\nu$.
Since $<m_{ee}>_{\rm min}$ is 
approximately equal to
$\sqrt{\Delta m_{23}^2}/\sin2\theta_{12}
=   ( 0.034  - 0.069 ) ~~\mbox{eV}$, it is
 consistent with recent experiments \cite{klapdor1,wmap}.

\section{$\mu\to e\gamma$ constraint}
The $Z_2$ even neutral components of the Higgs fields $\phi_I$ have 
tree-level FCNC couplings in the lepton  sector
as we can see from (\ref{Yue}).
Since the Yukawa couplings $y_2, y_4$ and $y_5$ are related to the
charged lepton masses (see (\ref{m2-m5})), they are very small.
As a consequence, the tree-level FCNCs are well suppressed
in the present model, as it is shown in ref. \cite{kubo4}.
The one-loop $\mu\to e\gamma$ amplitude
mediated by $\phi_I$ is negligibly small, too,
because of  the same reason.
In the following discussions, therefore, we consider
$\mu \to e \gamma$ mediated only by the $\eta$ exchange.

To this end, we  express the Yukawa couplings
in terms of the mass eigenstates.
We use the approximate unitary matrix (\ref{UeL})
for the charged leptons, and 
we find the Yukawa terms relevant for $\mu\to e\gamma$ are
given by
\be
{\cal L}_{Y^\nu} &=&
-Y^+_{ij} e_{Li} n_{j} \eta_+^{(+)}-
Y^- _{ij} e_{Li} n_{j} \eta_-^{(+)}
-Y^S_{ij}  e_{Li} n_{j} \eta_S^{(+)}+h.c.,
\label{Lh}
\ee
where
\be
Y^\pm &=& U_{eL}^T(Y^{\nu 1}\pm Y^{\nu 2})/\sqrt{2}~,~
Y^S=U_{eL}^TY^{\nu 3},
\ee
and
\be
Y^+ &\simeq &
\left(\begin{array}{ccc}
( h_{4}-2\epsilon_e h_2)/\sqrt{2} & 
h_{4}/\sqrt{2} &
0\\
h_2  + \epsilon_e h_{4}& 
 \epsilon_e  h_{4} &0 \\
0 & h_2 & 0
\end{array}\right),
\label{yukH}\\
Y^- &\simeq&
\left(\begin{array}{ccc}
h_{4}/\sqrt{2} &
( -h_{4}-2\epsilon_e h_2)/\sqrt{2}
 &0 \\
\epsilon_e h_{4} & 
h_2  - \epsilon_e h_{4} &0 \\
-h_2 & 0 & 0
\end{array}\right),
\label{yukm}\\
Y^S& \simeq &
\left(\begin{array}{ccc}
0& 0 & h_{3}\\
0 & 0&  \sqrt{2}\epsilon_e h_{3} \\
0 & 0 & 0
\end{array}\right)~,~
(\sqrt{2}\epsilon_e=m_e/m_\mu \simeq \sqrt{2}\sin\theta_{13}
\simeq 0.0048).
\label{yukL}
\ee
The inert Higgs fields $\eta_\pm, \eta_S$ 
and the Yukawa matrices $Y^\nu$'s are defined in (\ref{phipm})
and (\ref{Yun}), respectively.

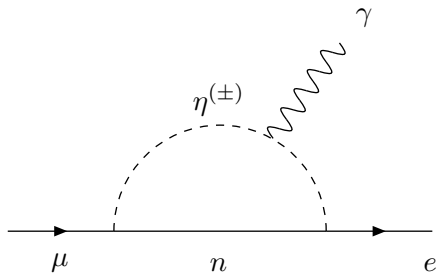
\begin{figure}[htb]
\begin{center}\begin{picture}(300,110)(0,45)
\ArrowLine(70,50)(110,50)
\Line(110,50)(190,50)
\ArrowLine(190,50)(230,50)
\Text(90,35)[b]{$\mu$}
\Text(230,35)[b]{$e$}
\Text(150,35)[b]{$n$}
\Text(150,100)[]{$\eta^{(\pm)}$}
\Text(205,131)[]{$\gamma$}
\Photon(195,121)(170,85){4}{5}
\DashCArc(150,50)(40,0,180){3}
\end{picture}\end{center}
\caption[]{\label{muegamma}One-loop diagram for
$\mu\to e\gamma$.}
\end{figure}
Using the Yukawa interaction term (\ref{Lh}),
we then compute the branching fraction 
of $\mu \to e \gamma$ from fig. \ref{muegamma}, and find
\be
B(\mu\to e\gamma)
&=&\frac{3\alpha}{64\pi G_F^2 } X^4
\simeq \left| X^2 900~\mbox{GeV}^2\right|^2,
 \label{mutoe}
\ee
where
\be
X^2 &\simeq &\frac{ h_4 h_2}{\sqrt{2 }}
\left[~\frac{F_2(M_1^2/m_+^2)}{m_+^2}-
\frac{F_2(M_1^2/m_-^2)}{m_-^2}~\right]
+h_3^2 ~\frac{m_e}{m_\mu}~ \frac{F_2(M_3^2/m_S^2)}{m_S^2},
\label{mueg}
\ee
and
\be
F_2(x) &=& {1-6x+3x^2+2x^3-6x^2\ln x \over 6(1-x)^4}.
\label{func-f2}
\ee
($m_\pm$ and $m_S$ are the masses of $\eta_\pm$ and $\eta_S$,
respectively, which are defined in (\ref{phipm}).)
We recall that $M_1=M_2$ because of the $D_6$ symmetry.
As we see the third term of (\ref{mueg}) contains
a  suppression factor $m_e/m_\mu$. 
There is no suppression factor
like $m_e/m_\mu$ for  the first two terms in  (\ref{mueg}),
but there are two cancellation
mechanisms, where these two mechanisms can be combined.
The first one takes place if  $M_1 >> m_+,m_-$,
for which
the leading terms in the parenthesis cancels, and one finds
\be
& &\frac{F_2(M_1^2/m_+^2)}{m_+^2}-
\frac{F_2(M_1^2/m_-^2)}{m_-^2}\nn\\
 & =&
-\frac{1}{M_1^4}
\left[~\frac{11}{6}(m_-^2-m_+^2)
-m_-^2 \ln (M_1^2/m_-^2)
+m_+^2 \ln (M_1^2/m_+^2)+O(1/M_1^2)~\right].
\label{f2-f2}
\ee
The second one is obvious because
of the relative sign in (\ref{f2-f2}):
it cancels exactly if $m_+=m_-$.
The second possibility appears naturally for 
relatively heavy $\eta_\pm$, because
$m_+^2-m_-^2\sim O(\lambda) v^2$ as we can see
from the Higgs potential (\ref{pot0}).
To get an idea on the size of the suppression
we compute the branching fraction
for $M_1=2$ TeV, $m_+ =700$ GeV and $m_-=750$ GeV,
and find
\be
B(\mu\to e\gamma) \simeq 1.6\times 10^{-12},
\ee
where we have neglected the contribution coming from
the third term in (\ref{mueg}) and used $|h_4 h_2|=1$.
This should be compared with 
$B(\mu\to e\gamma) \simeq 1.1 \times 10^{-9}$,
which one would obtain in the absence of the second term 
in the parenthesis.
As we see from (\ref{func-f2})  the function
$F_2(x)$ can vary between $1/12 (x=1)$ and $1/6 (x=0)$
for  $x \le 1$. From this we can roughly estimate the
lower bound of $m_S$ for the case that 
$n_S$ is a CDM candidate, i.e., $M_S < m_S$.
We find 
that $m_S > O(300)~\mbox{GeV}$ 
for $F_2 \simeq 1/12$ and
$h_3 \simeq 0.93$  should be satisfied to satisfy
$B(\mu\to e\gamma) < 1.2 \times 10^{-11}$.

Now, suppose that CDM is fermionic.
From Table I we see that only $D_6$ doublet and singlet right-handed
neutrinos, i.e. $n_I$ and $n_S$, come into   question.
If $n_I$ are CDM, then $M_1 <  m_\pm $ should be satisfied.
(Otherwise $n_I$  would decay into $\eta_\pm$.)
As we can find from (\ref{mueg}), 
we have to impose a fine tuning
of the form
\be
(m_+^2-m_-^2)/m_+^2 &\lsim& (m_+/500
\mbox{GeV})^2\times 10^{-2}
\ee
to sufficiently suppress $\mu\to e\gamma$ in this case.
This means that, for instance,  if $m_+=500$ GeV, 
then $m_-$ should be
within $500 \pm 5$ GeV.
Therefore, it is more natural to assume that
$M_1 > m_\pm$ to  suppress $\mu\to e\gamma$.
Then only the $D_6$ singlet right-handed neutrino $n_S$ remains
as a fermionic CDM candidate.
As we can see from the Yukawa matrices
(\ref{yukH}), (\ref{yukm}) and (\ref{yukL}),
only $\eta_S$ couples to  $n_S$ 
with $e_L$ and with $\mu_L$,
where the coupling with $\mu_L$ is suppressed
by  $m_e/m_\mu \simeq 0.005$.
Therefore, there would be  a clean signal if
the charged component of $\eta_S$ were produced at LHC.
In the next section we will investigate whether or not
the $\mu\to e \gamma$ constraint above
is consistent with the observed dark matter relic density
$\Omega_d h^2\simeq 0.12$ \cite{wmap}
assuming that $n_S$ is  CDM.

\section{Cold Dark matter}
Here we would like to investigate whether or nor
 $n_S$  can be a good CDM candidate.
For simplicity we assume  that
$m_S\simeq m_S^R \simeq m_S^I >>  m_S^{RI}$ (which
are defined in (\ref{lmeta})),
that is, the breaking of 
$U(1)_L'$ by the VEV (\ref{VEVphi}) is small
(which is needed to obtain small neutrino masses).
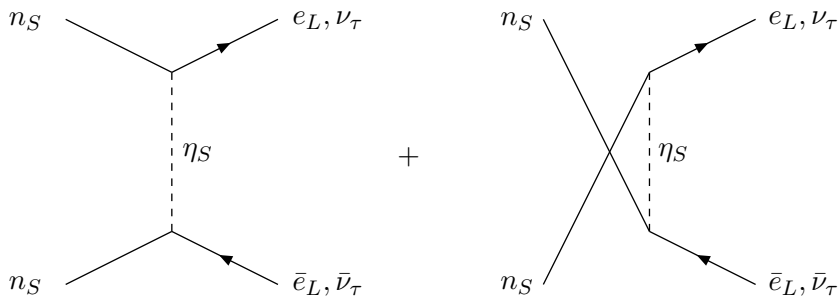
\begin{figure}[htb]
\begin{center}\begin{picture}(300,110)(0,45)
\Text(180,95)[b]{$+$}
\DashLine(90,130)(90,70){3}
\Line(50,150)(90,130)
\Text(35,150)[]{$n_S$}
\ArrowLine(90,130)(130,150)
\Text(136,150)[l]{$e_L, \nu_\tau$}
\Line(50,50)(90,70)
\Text(35,50)[]{$n_S$}
\ArrowLine(130,50)(90,70)
\Text(136,50)[l]{$\bar{e}_L, \bar{\nu}_\tau$}
\DashLine(270,130)(270,70){3}
\ArrowLine(270,130)(310,150)
\Text(316,150)[l]{$e_L, \nu_\tau $}
\ArrowLine(310,50)(270,70)
\Text(316,50)[l]{$\bar{e}_L, \bar{\nu}_\tau$}
\Line(230,150)(270,70)
\Text(221,150)[]{$n_S$}
\Line(230,50)(270,130)
\Text(221,50)[]{$n_S$}
\Text(100,100)[]{$\eta_S$}
\Text(280,100)[]{$\eta_S$}
\end{picture}\end{center}
\caption[]{\label{annihilation} Annihilation  diagram of
$n_S$.}
\end{figure}
To obtain a thermally averaged  cross section 
for the annihilation of two $n_S$'s, we
compute the relativistic cross section $\sigma$ 
from fig. \ref{annihilation} and  expand it
in powers of the relative velocity $v$ 
of incoming $n_S$ \cite{griest1}.
Note that the $\eta_\pm$ exchange
diagrams are suppressed,
which one sees from the Yukawa matrices (\ref{yukH})
and (\ref{yukm}). 
Lepton number violating diagrams being
proportional to $\Delta m_a^2$ in (\ref{deltama}) are also very small.
So they are annihilated mostly
into an $e^+-e^-$ pair and a $\nu_\tau-\bar{\nu}_\tau$ pair.

Using the result of ref. \cite{griest1}, we find 
in the limit of the vanishing lepton masses
\be
\sigma v &=& a + b v^2+\cdots,~
a =0, b = \frac{|h_3|^4 r^2 (1-2 r+2 r^2)}{24 \pi M_S^2},
\label{bandr}\\
  r &= & M_S^2/(m_S^2+M_S^2).
  \label{randy}
\ee
Then we can compute the relic density of $n_S$ from \cite{griest2}
\be
\Omega_d h^2 &=&\frac{Y_\infty s_0 M_S}{\rho_c/h^2}~
\mbox{with}~Y_\infty^{-1} =0.264 g_*^{1/2} M_{pl} M_S (3 b/x_f^2),
\label{omega}
\ee
where $Y_\infty$ is the asymptotic value of the ratio $n_{n_S}/s$,
$s_0=2970/\mbox{cm}^3$ is the entropy density at present,
$\rho_c=3 H^2/8 \pi G=
1.05 \times 10^{-5}h^2 ~\mbox{GeV}/\mbox{cm}^3$ is the critical density,
$h$ is the dimensionless Hubble parameter,
 $M_{pl}=1.22\times 10^{19}~ \mbox{GeV}$ is the  Planck energy,
and $g_*$ is the number of the effectively massless degrees of freedom
at the freeze-out temperature.
Further, $x_f$ is the ratio $ M_S/T$ at the 
freeze-out temperature and is given by \cite{griest2}
\be
x_f &=&\ln \frac{0.0764 M_{pl}(6 b/x_f) c (2+c) M_S}
{(g_* x_f)^{1/2} }.
\label{xf}
\ee
Using $g_*^{1/2}=10$ and $c=1/2$ \cite{griest2} we obtain
\be
\frac{ M_S}{\mbox{GeV}} &=&5.86\times 10^{-8}
 x^{-1/2}_f(\exp x_f ) 
\left[\frac{\Omega_d h^2}{0.12}\right],
\label{mn}\\
\frac{b } {\mbox{GeV}^2}&=&2.44 \times 10^{-11}
 x_f^2\left[\frac{0.12}{\Omega_d h^2}\right],
\label{meta}
\ee
where $b$ is  given in (\ref{bandr}). 

\begin{figure}[htb]
\includegraphics*[width=0.8\textwidth]{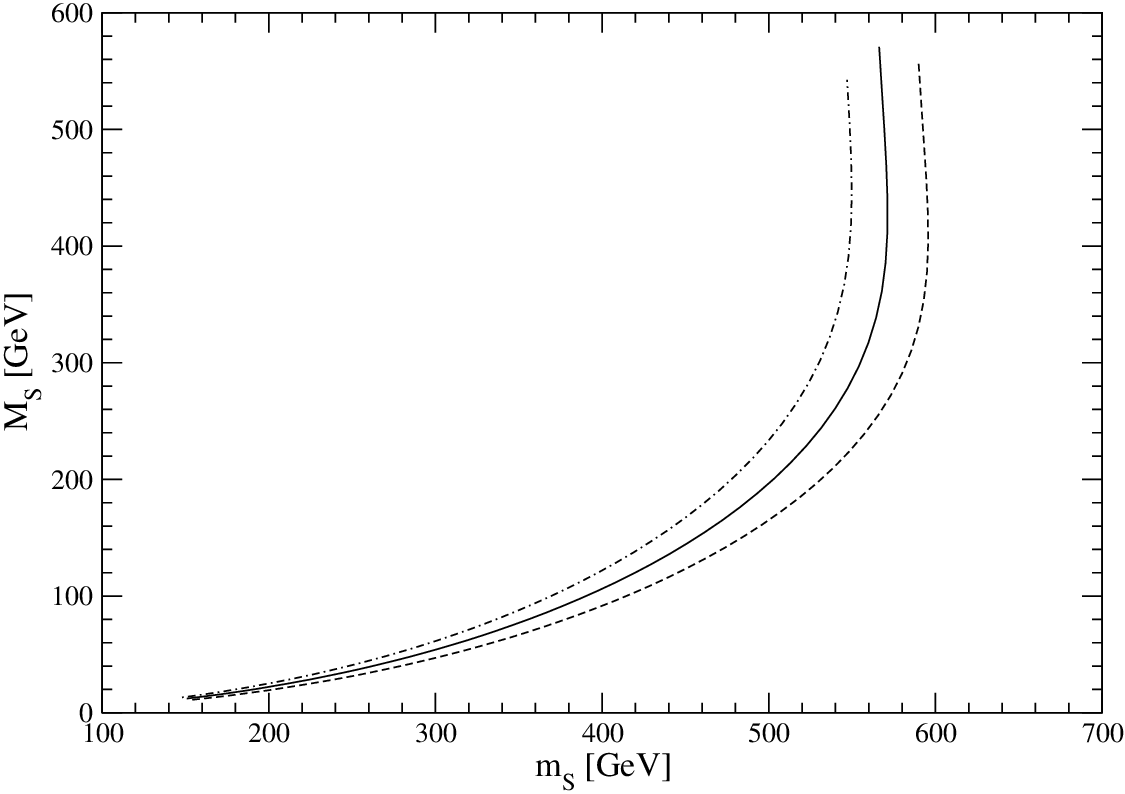}
\caption{\label{ms-ms-omega}\footnotesize
$M_S$ versus $m_S$ for $|h_3|=1.3$
and $\Omega_d h^2=0.13 (\mbox{dot-dashed}), 
0.12 (\mbox{solid})$ and $0.11(\mbox{dashed})$.
$M_S$ and $m_S$ denote the masses of
the CDM $n_S$ and the $D_6$ singlet neutral Higgs,
respectively.
}
\end{figure}

\begin{figure}[htb]
\includegraphics*[width=0.8\textwidth]{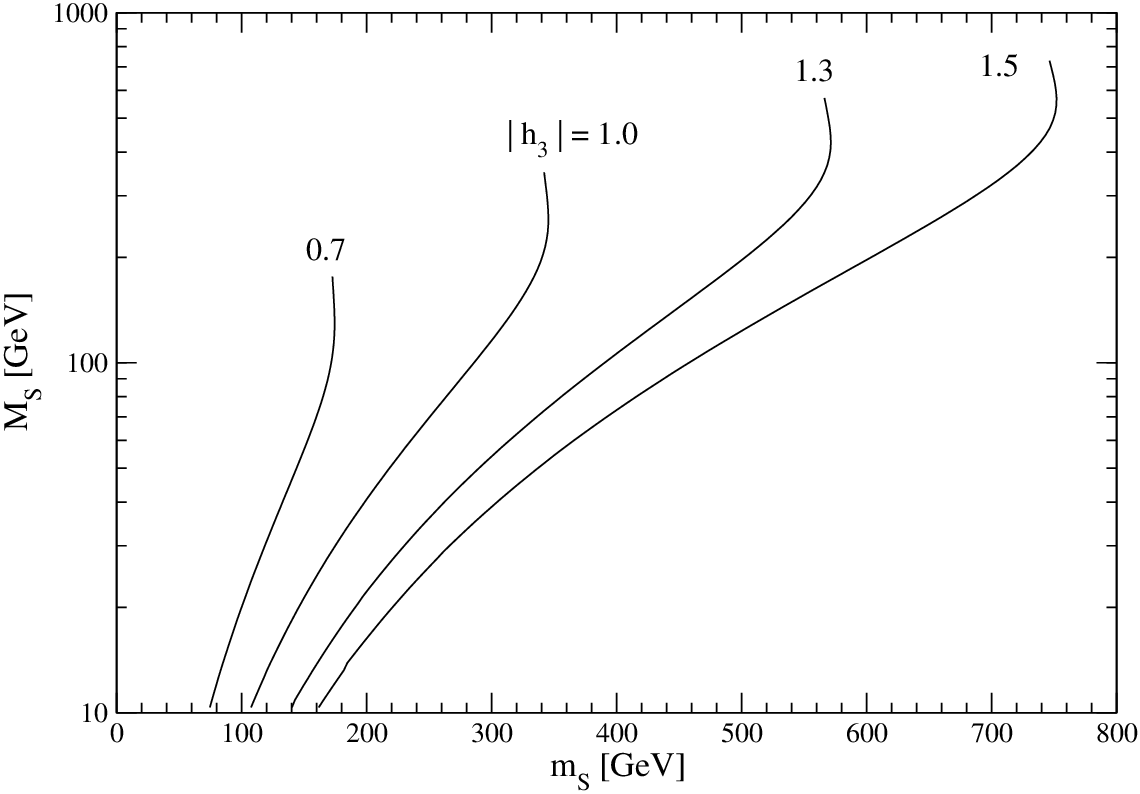}
\caption{\label{ms-ms-h3}\footnotesize
$M_S$ versus $m_S$ 
at  $\Omega_d h^2=0.12$ for $|h_3|=1.5, 1.3,
1.0, 0.7$.
$M_S < m_S$ should be satisfied
because $n_S$ is assumed to the CDM candidate.
}
\end{figure}

\begin{figure}[htb]
\includegraphics*[width=0.8\textwidth]{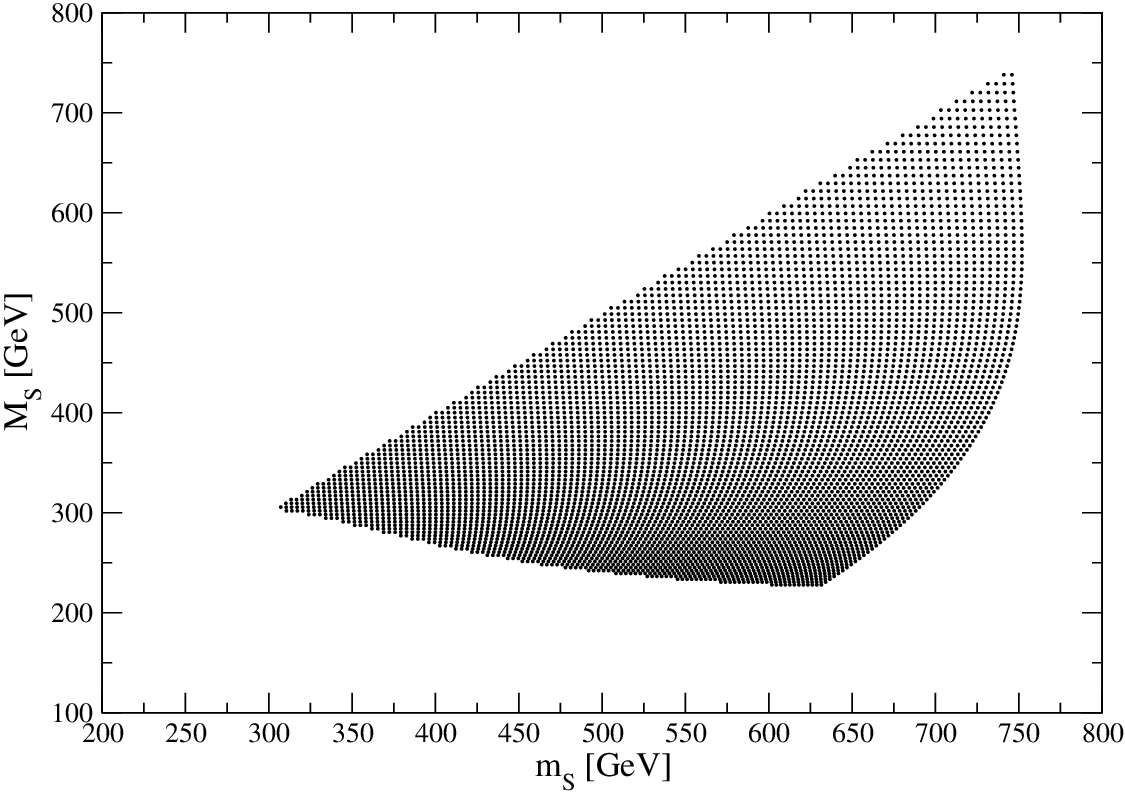}
\caption{\label{ms-ms}\footnotesize
The region in the $m_S-M_S$ plane
in which $\Omega_d h^2=0.12,
B(\mu\to e \gamma) < 1.2\times 10^{-11}$ and
$| h_3| <1.5$ are satisfied.
}
\end{figure}

Since $b/|h_3|^4$ is a
function of $ M_S$ and $m_S$, eqs. (\ref{mn}) and (\ref{meta})
give $ M_S$ and $m_S$ in unit of GeV for a given set
of $|h_3|^2, x_f$ and $\Omega_d h^2$. 
In fig. \ref{ms-ms-omega},
$M_S$  against $m_S$ is plotted
for $\Omega_d h^2=0.13  (\mbox{dot-dashed}), 
0.12 (\mbox{solid})$ and $0.11 (\mbox{dashed})$ \cite{wmap},
where $|h_3|$ is fixed at $1.3$.
We see from the figure that the dark matter mass $M_S$ 
and the $D_6$ singlet Higgs mass $m_S$ 
should be closely related with each other 
to obtain an observed relic  density  of
dark matter.
We plot in fig. \ref{ms-ms-h3},
$M_S$   against $m_S$
for $y =0.3, 0.5, 0.7,  1.0$.
As we can see also from the figure, the dark matter constraint
requires that the dark matter mass $M_S$  increases
as the Higgs mass $m_S$ increases and
reaches at its maximal value $m_S$ at a certain 
value of $m_S$.
We see that $m_S$ can not exceed $\sim 750$ 
GeV for the perturbative regime $|h_3| \leq 1.5$.
Therefore, it is by no means trivial to satisfy
the $\mu\to e\gamma$ constraint.
In fig. \ref{ms-ms} we present
the allowed region in the $m_S-M_S$ plane, in which
$\Omega_d h^2=0.12$
and $B(\mu\to e \gamma) < 1.2\times 10^{-11}$
are satisfied, where we assume $|h_3| <1.5$
and only the last term in $X^2$ of (\ref{mueg})
contributes to $\mu\to e\gamma$.
If we allow larger $|h_3|$,
then the region expands  to
 larger $m_S$ and $M_S$,
 and for  $|h_3|\lsim 0.8$ there is no allowed region.
 As we can also see from fig. \ref{ms-ms},
the mass of the CDM and the mass of the inert Higgs should be larger than 
about $230$ and $300$ GeV, respectively.
If we restrict ourselves to
a perturbative regime, they should be lighter than about $750$ GeV.
 
\section{Conclusion}
We have assumed that two important issues, the origin of neutrino masses and 
the nature of CDM, are related,
and considered
 a non-supersymmetric extension
of the standard model with a family symmetry based on
$D_6 \times \hat{ Z}_2 \times Z_2$.
The gross structure of the Yukawa couplings is
fixed by $D_6$, while $\hat{ Z}_2$ is responsible to
suppress FCNCs in the quark sector, e.g. in
the mixing of the neutral meson systems.
They are spontaneously broken together with $SU(2)_L\times U(1)_Y$.
The remaining $Z_2$ is unbroken; 
it  forbids the tree-level neutrino masses
and simultaneously ensures the stability of CDM 
candidates.
From the assumption that CDM is fermionic
we can single out the $D_6$ singlet right-handed neutrino
as the best CDM candidate.
We find that the $D_6$ singlet  inert charged Higgs
with a mass between $300$ and $750$ GeV
 has a tree-level coupling to 
the electron, the muon and the tau 
with the relative strength $(1, m_e/m_\mu,0)$, respectively.
The lower value results from the $\mu \to e \gamma$ constraint,
while we obtain the upper value from the assumption 
that the Yukawa sector remains within a framework
of perturbation theory.
(This assumption is refelected on the condition $M_S < m_S$.)
From this observation we have concluded that 
the charged Higgs decays mostly into
an electron (or a positron) with a large missing energy, where
the missing energy is carried away by the CDM candidate.
This will be a clean signal at LHC.

\vspace{0.5cm}
\noindent
{\large \bf Acknowledgments}\\
We would like to thank E.~Ma and D.~Suematsu
for useful discussions. 
This work is supported by the Grants-in-Aid for Scientific Research 
from the Japan Society for the Promotion of Science
(\# 18540257).
                                                                           
\newpage
\bibliographystyle{unsrt}

\end{document}